\documentclass[sn-mathphys,Numbered]{sn-jnl}

\usepackage{graphicx}%
\usepackage{multirow}%
\usepackage{amsmath,amssymb,amsfonts}%
\usepackage{amsthm}%
\usepackage{mathrsfs}%
\usepackage[title]{appendix}%
\usepackage{xcolor}%
\usepackage{textcomp}%
\usepackage{manyfoot}%
\usepackage{booktabs}%
\usepackage{algorithm}%
\usepackage{algorithmicx}%
\usepackage{algpseudocode}%
\usepackage{listings}%
\usepackage{subcaption}
\usepackage{caption}

\theoremstyle{thmstyleone}%
\theoremstyle{thmstyletwo}%

\theoremstyle{thmstylethree}%

\begin{document}

\title[Article Title]{FLoW3 - Web3 Empowered Federated Learning}


\author[1]{\fnm{Venkata Raghava} \sur{Kurada}}\email{venkataraghavakurad@sssihl.edu.in}

\author[2]{\fnm{Pallav Kumar} \sur{Baruah}*}\email{pkbaruah@sssihl.edu.in}

\affil[1,2]{\orgdiv{Department of Mathematics and Computer Science}, \orgname{Sri Sathya Sai Institute of Higher Learning}, \orgaddress{\street{Street}, \city{Puttaparthi}, \postcode{515134}, \state{Andhrapradesh}, \country{India}}}

\abstract{Federated Learning is susceptible to various kinds of attacks like Data Poisoning, Model Poisoning and Man in the Middle attack. We perceive Federated Learning as a  hierarchical structure, a federation of nodes with validators as the head. The process of validation is done through consensus by employing Novelty Detection and Snowball protocol, to identify valuable and relevant updates while filtering out potentially malicious or irrelevant updates, thus preventing Model Poisoning attacks. The opinion of the validators is recorded in  blockchain and trust score is calculated. In case of lack of consensus, trust score is used to determine the impact of validators on the global model. A hyperparameter is introduced to guide the model generation process, either to rely on consensus or on trust score.  This approach ensures transparency and reliability in the aggregation process and allows the global model to benefit from insights of most trusted nodes. In the training phase, the combination of IPFS , PGP encryption provides : a) secure and decentralized storage  b) mitigates single point of failure making this system reliable and c) resilient against man in the middle attack.  The system is realized by implementing in python  and Foundry for smart contract development. Global Model is tested against data poisoning by flipping the labels and by introducing malicious nodes. Results found to be similar to that of  Flower.
}

\keywords{Federated Learning, Novelty Detection, Snowball Consensus Protocol, Web3  Technology}



\maketitle

\section{Introduction}\label{sec1}
There is a steep increase in the usage of the various wearable devices, spanning from smartphones to smartwatches, that  generates data which is quite sensitive and private in nature. This data can be used to train machine learning models and improve the performance. Transferring this data, over the internet, to a datacenter raises concerns about data security and privacy.  This has driven the advent of  Federated Learning\cite{r1}. Federated Learning faces challenges from various attacks like Man in the middle attack, lack of transparency and decision making power to the participating entities.

There are stringent data protection laws like GDPR \cite{r22} , Digital CCPA\cite{r23}, DPDP Act\cite{r36}, that enforces a wide variety of restrictions while processing and storing of the data. The synergy of the AI and Federated Learning is a promising path to develop privacy centric machine learning models and enables models to leverage on data while preserving the privacy. In this paper, we demonstrate a way to fulfil this promise.

Federated Learning, is a machine learning paradigm where all the nodes are arranged in a star topology with a central server and various edge devices (nodes) connected to it. The central server will send the initial global model to all the nodes. Upon receiving this model, the nodes will train the model on the data that is locally available with the on-device hardware. After training the model, the nodes send the model updates to the central server. The central server will then implement a Federated Aggregation algorithm, thus creating a new global model. This process repeats until the model converges.

There are works \cite{r2}\cite{r3} demonstrating the feasibly preserving data privacy in federated learning using security measures like Differential Privacy \cite{r37}, Homomorphic Encryption \cite{r34}
Federated learning is widely used in different domains ranging from health care  to image classification on unlabelled data \cite{r4} \cite{r5} \cite{r35}. Often times this type of setup is controlled by a single organization in an isolated environment. 
There is a need for the decentralized platform, one that can effectively enable Federated Learning to be adopted widely. This platform should empower participants with  a transparent decision-making process and incentivise the participants for the active involvement of the users.

    Web3 \cite{r6} technology is a suite of distributed technologies like Blockchain \cite{r7}, Inter Planetary File System (IPFS) \cite{r8}, Cryptocurrencies, Smart Contract\cite{r17}. Instead of relying on central authorities to control and manage user data, Web3 aims to distribute control among a network of participants. It envisions the process of giving control back to the user. The blockchain ensures transparency and trust, while storage is facilitated by various distributed storage built on IPFS. This suite of technology relies on  distributed computing capability of its nodes to cater its computation needs.

Snowball Protocol\cite{r9} is a leaderless, probabilistic consensus protocol that is scalable in nature. It is efficient in terms of communication, where each node is required to communicate with a constant number of nodes, contributing to its low overhead. These attributes make Snowball protocol a suitable choice for the proposed platform.
    
Main contributions of this work are:

\begin{itemize}
  \item Design and implementation of federated learning system in decentralized way by introducing a hierarchy between the nodes

  \item Considering Federated Learning from a consensus prospect, bringing transparency in the decision-making process.

  \item Blockchain based trust calculation solution as a foolproof mechanism to generate the model.
\end{itemize}

\section{Literature Review}
Federated learning was initially proposed by McMahain et al\cite{r1}. Konečný et al\cite{r38} proposed a set of optimizations on representation of weights to reduce the communication in federated learning setup. 
The process of Federated Learning is susceptible to Data Poisoning Attack, Gradient Leakage Attacks, as discussed in \cite{r10} \cite{r11}. 
In order to make the federated learning a successful participatory platform, we need to have a profitable design. Xu et.al\cite{r12} proposes a time based auction procedure  that will be economically profitable to both users and platform owners. 

While \cite{r13}\cite{r14}\cite{r15} proposes to use the blockchain in order to  tackle these attacks, Kim et al\cite{r16} proposed BlockFL architecture where the model updates were exchanged and verified, the work also performs the end to end latency model of BlockFL.

Differential Privacy introduces carefully calibrated noise to individual data points before aggregation on the other hand, Homomorphic encryption enables the required computations to be performed on the encrypted data, thus enabling the compliances with the mentioned acts.
In this work, we primarily focus on designing a platform that relies on Web3 technologies to implement Federated Learning in a transparent and hierarichal manner with privacy as a key factor. 

\subsection{Blockhain}
Blockchain\cite{r7} was initially proposed by Satoshi Nakamoto. It is a decentralized and distributed system that is immutable in nature. Blockchain is constituted by blocks,  which stores a set of transactions. Once a block is full, the hash of the current block is stored in its successive block. If a content of a previous block is modified, the change will be propagated via hash through all blocks, thus facilitating a tamper proof mechanism.

    Ethereum\cite{r17} with Ether as its native token, is one of the popular public blockchain networks that was initially proposed by Valtamir Butanik. With the introduction of the Smart Contract\cite{r17}, the Ethereum blockchain has transformed into a Turing complete machine. These Smart Contracts enable the user to, programmatically, create a set of rules that can be executed on the blockchain. Smart contracts, developed using Solidity, are utilized in various fields such as supply chain management \cite{r31} and finance \cite{r30} to streamline processes and ensure secure and transparent transactions.
\subsection{Novelty Detection}
Novelty detection\cite{r18} refers to the process of identification of abnormal patterns that are present in huge amounts of normal data. In this case, the emphasis is more on detecting whether an unseen observation is an outlier or a novelty. Furthermore, the main idea is to learn the definition of the "normality" by training a model on multiple positive instances that are not anomalies.  Based on the prediction of this trained model, a score is given to an unseen observation. This score is then compared with the decision threshold, thus determining the novelty of the observation.
According to Marco A.F. Pimente \cite{r18} the novelty detection can be classified in the following types:
\begin{itemize}
\item Probabilistic Novelty Detection: A type of novelty detection algorithms that are based on the estimation of the generative probability density function of the data. A threshold is imposed on the  resultant distribution for novelty. Based on which, a given point is classified as novelty or anomaly. 
\item Distance Based Novelty Detection: 
In this type of algorithms, novelty of a particular record is identified based on the distance by using  various clustering algorithms. Greater the distance from its neighbours, higher the chances of being a anomaly. 
\item Reconstruction Based Novelty Detection: 
In this type of novelty detection, a model of the given normal data is created. This model is later used to create or reconstruct new instances of data. The difficulty of reconstructing a new data record is considered as metric.  Higher the difficulty to generate the data point, greater the chances of considering it as anomaly. Auto encoders and PCA are widely used in this type of novelty detection
\end{itemize}
The proposed platform uses Local Outlier Factor\cite{r19} algorithm for novelty detection.

\subsection{Snowball Protocol}

    Snowball Protocol\cite{r9}, is a leaderless Byzantine fault tolerance protocol from the Snow family. Snowball is a single decree binary consensus protocol, that is highly scalable and probabilistic in nature. This consensus protocol  is based on the meta stability and network subsampling, i.e. repeatedly polling a random subset of participants to attain consensus. This concept of network subsampling perturbs the conflicting, thus causing one of the decisions to gain advantage over another decision. Unlike traditional consensus protocol,  there is no need for each node to acquire the accurate knowledge of the participant of the decision-making process. A modified version of the snowball protocol in synergy with Direct Acyclic graph is used in Avalanche\cite{r20} blockchain as a consensus mechanism. The Snow Family of protocols are quiescent and energy efficient as they do not require constant participation of nodes when there is no decision-making process, unlike Proof of Work \cite{r7}.
    
\begin{algorithm}[h!tbp]
\caption{Snowball Protocol}
\begin{algorithmic}[1]
\Function{Snowball}{}
    \State $\text{decision1} \gets 0, \text{decision2} \gets 1$
    \State $\text{prev\_decision} \gets 0, \text{count} \gets 0$
    \State $\text{confidence\_counter} \gets [0, 0]$
    \State $\text{decided} \gets \text{false}$
    \State $\text{final\_decision} \gets \text{NONE}$
    \While{$\text{decided} \neq \text{true}$}
        \State $\text{opinion} \gets \text{get opinion from } k \text{ peers}$
        \State $\text{confidence\_counter}[0] \gets \text{opinion.count}(\text{decision1})$
        \State $\text{confidence\_counter}[1] \gets \text{opinion.count}(\text{decision2})$
        \State $\text{max\_val} \gets \text{max}(\text{confidence\_counter})$
        \State $\text{max\_ind} \gets \text{confidence\_counter.index}(\text{max\_val})$
        \If{$\text{max\_val} > \alpha$}
            \If{$\text{max\_ind} == \text{previous\_decision}$}
                \State $\text{count} \gets \text{count} + 1$
            \Else
                \State $\text{previous\_decision} \gets \text{max\_ind}$
                \State $\text{count} \gets 0$
            \EndIf
            \If{$\text{count} \geq \beta$}
                \State $\text{decided} \gets \text{true}$
                \State $\text{final\_decision} \gets \text{max\_ind}$
            \EndIf
        \EndIf
    \EndWhile
\EndFunction
\end{algorithmic}
\end{algorithm}
Snowball Protocol (Algorithm 1) has the following parameters.
\begin{itemize}
    \item Number of Nodes(N): Total number of nodes in the network that are participating in the consensus.
    \item Quorum Size(k): Number of nodes that are subsampled during decision-making.
    \item Alpha($\alpha$): Minimum number of nodes required to decide the majority in a quorum
    \item Beta($\beta$): Minimum number of consecutive rounds in favour of a decision required to terminate the consensus protocol. 
\end{itemize}

A node taking part in the decision-making process starts in an undecided state. Then the node proceeds to initiate a query to k nodes, thus initiating the query process. Upon receiving a query, the undecided node will respond with the state it received and initiate its own query. On the other hand, a decided node will simply respond with its decision. After receiving k responses, a node checks whether majority is found or not, using $\alpha$. If majority is found, then the node updates the confidence  counter with respect to that specific decision. If the $\alpha$ fraction is not met, then the confidence counter for the specific decision is reset to zero. This process continues until one of the confidence counters reaches $\beta$.

\subsection{Federated Learning}
Federated learning\cite{r1}, is a machine learning paradigm where the aim is to create a centralized model without transmitting data to a centralized server. Unlike conventional machine learning setting, where the data is transmitted to a centralized server and the model training takes place in a highly controlled environment, Federated Learning facilitates on device training of the model.

    Federated Learning is arranged in a star-shaped setting, where the centralized server is at the top, followed by the clients that train the actual model. To enhance the privacy, instead of sending actual model weights to the centralized server, only the model updates were sent to the centralized server. The centralized server implements an aggregation algorithm, generally Federation Average Algorithm, where the aggregation of the weights under a federation is done. 
Based on the type of data a particular federated learning deals with, FL can be classified into two types:
\begin{itemize}
    \item Horizontal Federated Learning: This type of federated learning is used for the datasets which have similar features but different values. For example, “items bought” of customer from two different stores.
    \item Vertical Federated Learning: This type of setting is used for the data sets which have overlapping features of the same instances. For example, data of a particular user base in two different organizations
\end{itemize}

\subsection{IPFS}
Interplanetary Planetary File System(IPFS) \cite{r8} is a peer to peer distributed file system that aims to connect all the nodes  under the same file system. This file system makes use of Merkel DAG to create a versioned file system and to verify the integrity of the files. All the files in the IPFS are addressed using unique Content Identifiers(CIDs). CIDs are constituted by the hash function and a multicoded representing the version of the CID. SHA-1, SHA-256 or BLAKE2 are commonly used as hash functions.
    
    As soon as a node in IPFS accesses a file or hypermedia content, it caches the files that it downloaded, thus making it available to the other nodes in the network. As the storage in these nodes is quite limited, this cache of the nodes must be cleared periodically, this process is referred to as IPFS garbage collection. Pinning is the process that ensures the data to be accessible and prevents this garbage collection.
\section{System Design And Architecture}

In this section, we will provide an overview of the system architecture for the proposed system, highlighting the key components and their interactions and overall flow of the proposed model.

\subsection{System Design}
\begin{figure}
    \centering
    \includegraphics[width=0.5\linewidth]{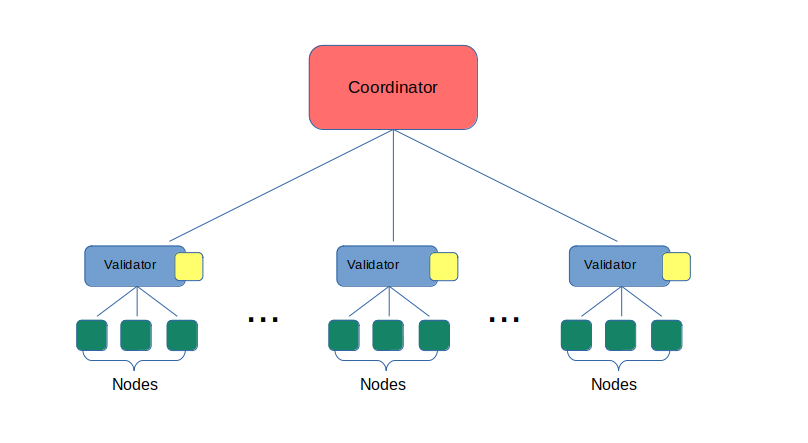}
    \caption{Architecture of Proposed Platform}
    \label{fig1}
\end{figure}

System design forms the foundation of a framework and aids in the development of the system from its theory. The proposed system has 3 different roles namely Node, Validator, Decentralized Aggregator arranged in hierarchical structure as depicted in Figure \ref{fig1}. These are further explained in the following section.
\subsubsection{Node}
It is the basic building block of the proposed architecture. Nodes are used to train the model with the data that is available locally. Nodes send the encrypted model updates to Validators via IPFS.

\subsubsection{Validator}
Validator is a specialized node that oversees  the process of federated learning of the nodes in its federation. The validator computes  a weighted average of the model updates, weighted by the number of examples each node used for training, thus creating the model. After the creation of the model, the validator then trains a novelty detection model with the weights that are submitted by the nodes. Later, the validator will participate in the Snowball protocol to attain the consensus on the validity of the weights submitted by the other validators.

\subsubsection{ Decentralized Aggregator}
It is a specialized node responsible for generating of the global model. The model generation can happen either by implementing the federated average algorithm on accepted weights or by implementing weighted average, weighted on influence derived from the blockchain. This node also initialises Snowball protocol, where the validation of proposed weights occurs. From a conventional federated learning context, a decentralized aggregator can be viewed as a Centralized Server. 

\subsection{System Flow}
\begin{figure}[t]
    \centering
    \includegraphics[width=0.5\linewidth]{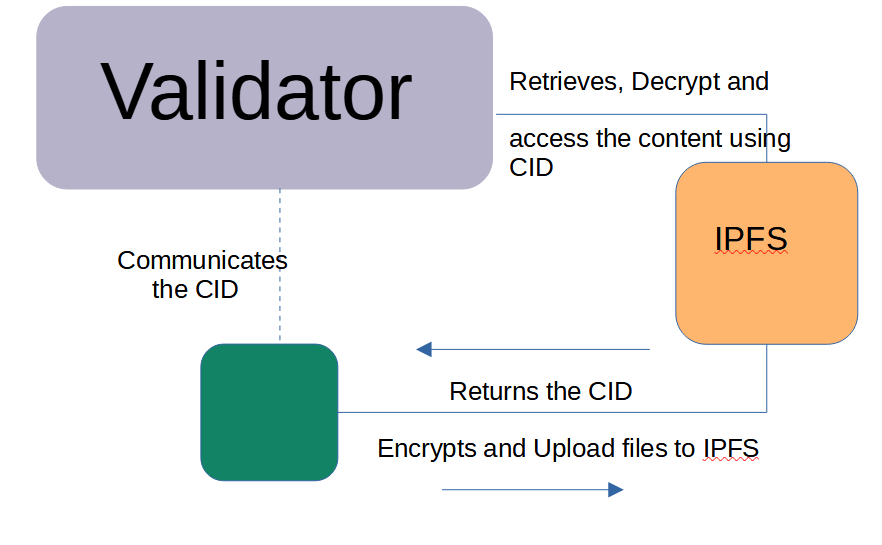}
    \caption{Interaction With IPFS}
    \label{fig2}
\end{figure}

 Once the hierarchy, constituted by Decentralized Aggregator, Validators, and Nodes is set up, the Decentralized Aggregator will invoke the smart contract and collect surety from Validators. The validator will initialize the federated learning process  for its federation of nodes. After the completion of training in the node, the weights are encrypted using the Pretty Good Privacy (PGP)\cite{r21} encryption algorithm and uploaded to IPFS. The CID of the uploaded file is then sent to the validator. The validator can access the IPFS, retrieve the encrypted file, decrypt it, and obtain the decrypted weights. The introduction of IPFS is done to prevent the man in the middle attack as depicted in Figure \ref{fig2}. 

At validator level, these weights were used a model using federated average algorithm. Additionally, the validators also train their own novelty detection models based on these weights. These novelty detection models will enable the validators in identifying new or unusual patterns.

    The weights of the model generated were then broadcast to all the validators, where the Snowball protocol is used to attain consensus. The main goal of this decision-making process is to determine the validity of the model generated by a particular validator. If all the nodes reach a consensus for a specific set of weights, then it is used in generation of the global model. The opinion of each validator is recorded in the blockchain.
    
Sometimes it might happen that, by the end of the Snowball protocol, none of the proposed weights were accepted. In this case, the global model is generated using weighted average weighted on influence of a particular validator. Calculation of the trust score is an important step in finding out influence. The Algorithm \ref{alg2} is used to determine the trust score. 
    
\begin{algorithm}
\caption{Trust Calculation}\label{alg2}
\begin{algorithmic}[1]
\Function{trust\_calculation}{$\text{final\_opinion}, \text{individual\_opinion}$}
    \State $\text{trust\_score} \gets 0, \text{count} \gets 0$
    \State $\text{total} \gets \text{len}(\text{final\_opinion})$
    \While{$\text{count} < \text{total}$}
        \If{$\text{final\_opinion}[\text{count}] == \text{individual\_opinion}[\text{count}]$}
            \State $\text{trust\_score} \gets \text{trust\_score} + 10$
        \Else
            \State $\text{trust\_score} \gets \text{trust\_score} - 10$
        \EndIf
        \State $\text{count} \gets \text{count} + 1$
    \EndWhile
    \State \textbf{return} $\max(\text{trust\_score}, 1)$
\EndFunction
\end{algorithmic}
\end{algorithm}

\begin{algorithm}
\caption{Blockchain Based Validator Influence Calculation}
\label{alg:blockchain_model_generation}
\begin{algorithmic}[1]
\State Let $O_1, O_2, \ldots, O_n$ be the opinions of $V_1, V_2, \ldots, V_n$ with respect to $W_1, W_2, \ldots, W_n$.
\State Let $w_1, w_2, \ldots, w_x$ be the accepted weights.
\State Let $C$ be the opinion obtained by consensus.
\Function{Validator\_Influence\_Calculation}{$C$}
    \State $opinions \gets [O_1, O_2, \ldots, O_n]$
    \State $accepted\_weights \gets [w_1, w_2, \ldots, w_x]$
    \State $trust \gets []$
    \State $residual\_trust \gets []$
    \State $max\_influence \gets \frac{1}{\zeta}$
    \State $shared\_influence \gets 1 - (accepted\_weights.\text{length} \times max\_influence)$
    \For{$y$ \textbf{in} $opinions$}
        \State $t \gets \text{trust\_calculation}(C, y)$
        \State $trust.\text{add}(t)$
    \EndFor
    \State $influence \gets \text{standardization}(trust)$
    \For{$y$ \textbf{in} $accepted\_weights$}
        \State $influence[\text{weights.index}[y]] \gets \text{max\_influence}$
    \EndFor
    \For{$y$ \textbf{in} $trust$}
        \State $t1 \gets y \times \text{residual\_influence}$
        \State $\text{shared\_influence}.\text{add}(t1)$
    \EndFor
    \State $\text{final\_influence} \gets \text{influence} + \text{shared\_influence}$
\EndFunction
\end{algorithmic}
\end{algorithm}
The function takes two arguments, \textit{final\_opinion} which corresponds to the final decision that is made by all the nodes after implementing Snowball protocol, the second argument corresponds to \textit{individual\_opinion} of each validator. This argument is obtained from the blockchain. The opinion of each validator is compared to the final consensus of the whole network on a particular weight. If both are the same, the trust score is incremented, else it is decremented. Since the weights in weighted average cannot be negative, instead of negative trust score, a minimum value is returned. 
\begin{figure}
\centering
\begin{minipage}{0.8\textwidth}
\hrulefill

\textbf{Input:}
\begin{itemize}
  \item Let \(W_1, W_2, \ldots, W_n\): Weights proposed by corresponding Validators \(V_1, V_2, \ldots, V_n\) respectively.
  \item \(x\): Number of accepted weights obtained from Snowball Protocol.
  \item \(\zeta\): Threshold value set to \(i\).
\end{itemize}

\textbf{Output:} Generated model

\textbf{Algorithm:}
\begin{lstlisting}[mathescape=true]
if (x >= $\zeta$) 
{
  Generate final model by Averaging the Accepted Weights;
} 
else 
{
  Generate final model based on 
  Blockchain Based Validator Influence;
}
\end{lstlisting}
\hrulefill
\caption{Algorithm for Model Generation}
\end{minipage}
\end{figure}
In order to offer a better control on the process of generating the global model, we introduce a new parameter called \textit{Minimal Consensus Index}($\zeta$). Minimal Consensus Index corresponds to the minimum number of accepted weights required to generate the global model using the Federated Averaging Algorithm. If the number of accepted weights is less than the $\zeta$ then the global model is generated based on the trust score of the validators. The value of the $\zeta$ should lie between 0 and n, where n is the total number of validators. The higher the value of ($\zeta$), the greater the chances of generating the model based on validator trust score.  

Based on the trust score and ($\zeta$) value, the individual influence is then calculated as follows.
\begin{itemize}
    \item Trust Score is standardized to obtain the influence.
    \item \textit{Maximum Influence}, the inverse of $\zeta$, is awarded to all the accepted weights.
    \item The \textit{Shared Influence} is then distributed evenly among all validators.
    \item \textit{Final Influence} is obtained by summing both the \textit{Maximum Influence} and the shared influence.
    \item A Global Model is generated using a weighted average weighted by the \textit{Final Influence}.
\end{itemize}
Algorithm 3 depicts the influence calculation workflow. The final system flow of the proposed platform is depicted in the Fig 3.

\subsubsection{Example}
Let C be the coordinator and V1,V2,V3,V4,V5 be the validators, let $\zeta$ be 5. After completion of training in node, the validator upon receiving model updates, implement Federated Average Algorithm and generates the weights of the model. Let W1,W2,W3,W4 and W5 be the weights that are generated by V1,V2,V3,V4,V5 respectively. Then the Snowball protocol is invoked. Table \ref{tab1} depicts the opinion of the validators with respect to proposed weights. Where, 1 represents the acceptance of the particular weight, and 0 represents the rejection.  It is evident that W1 and W5 are the accepted weights, as all the validators have approved these weights.

\begin{table}
\caption{Opinion of the Validators}\label{tab1}
\begin{tabular}{@{}llllll@{}}
\toprule
Validator& W1\footnotemark[1]  & W2 & W3 & W4 & W5\footnotemark[1]\\
\midrule
V1 & 1 & 0 & 0 & 1 & 1   \\
V2 & 1 & 0 & 1 & 1 & 1 \\
V3 & 1 & 1 & 0 & 0 & 1 \\
V4 & 1 & 0 & 0 & 0 & 1 \\
V5 & 1 & 1 & 0 & 0 & 1 \\
C & 1 & 0 & 0 & 0 & 1 \\ 

\botrule
\end{tabular}
\footnotetext[1]{Accepted Weights} 
\end{table}

If $\zeta$ is set to 5. The final model is generated using  Validator influence as depicted in Table \ref{tab2} 1. Trust score is normalized, as $\zeta$ is set to 5 the maximum impact of  each validator will be 0.2. Since  W1 and W5 were accepted, they were awarded  a weightage of 0.2. All the validators will share the remaining influence evenly among them. The sum of these two will yield the final validator influence, which is used as weight in the weighted average while generating the final model. The calculations are depicted in \ref{tab2}
If $\zeta$ is set to 2 or 1 then the final model is generated by averaging the accepted weights.

\begin{table}
\caption{Opinion of the Validators}\label{tab2}
\begin{tabular}{@{}llllll@{}}
\toprule
Title & W1  & W2 & W3 & W4 & W5\\
\midrule
Trust Score&30&10&30&50&30 \\
Influence&0.2&0.07&0.2&0.33&0.2 \\
Residual Influence(for 0.6)&0.12&0.04&0.12&0.20&0.12 \\
Final Influence&0.32&0.04&0.12&0.20&0.32\\
\botrule
\end{tabular}
\end{table}

\section{Opinion Collector}
Opinion Collector is a smart contract used in the proposed platform.This contract was constituted by getter and setter methods and used to maintain the records in a much more transparent and distributed manner. One of the key motivations to implement this smart  contract is to conduct the decision-making process as transparently as possible. Most of the functions in this smart contract can only be invoked by the coordinator, this restriction placed to ensure the authenticity of the data in Smart Contract

\begin{algorithm}
\caption{Opinion Collector}
\label{opinoncollector}
\caption{Algorithm for OpinionCollector}
\begin{algorithmic}
\Function{addValidator}{$\text{address}$}
    \If{$\text{msg.value} > 0.005$}
        \State add it to list of validators
        \State \textbf{return} (success)
    \Else
        \State \textbf{return} (failure)
    \EndIf
\EndFunction

\Function{setup}{}
    \If{$\text{msg.sender} == \text{coordinator}$}
        \State Initialize mapping to store opinion called \textit{store} with validator number of keys
    \EndIf
\EndFunction

\Function{get\_opinion}{}
    \If{$\text{msg.sender} == \text{coordinator}$}
        \State \textbf{return} \textit{store}
    \EndIf
\EndFunction

\Function{set\_opinion}{$\text{key}, \text{opinion}$}
    \If{$\text{msg.sender is validator or} \quad \text{msg.sender} == \text{coordinator}$}
        \State \textit{store}[$\text{key}$] $\gets \text{opinion}$
    \EndIf
\EndFunction
\end{algorithmic}
\end{algorithm}

Algorithm \ref{opinoncollector} explains the functioning of the smart contract for the proposed system. It has 4 functions that are used for implementing blockchain based trust. Most of these functions are invoked by the coordinator to verify the trustworthiness of the validators. This smart contract uses the following data structures:
\begin{itemize}
    \item An array is used to keep track of the validators, by holding their public key,  \textit{valid\_validators}.
    \item A mapping which has a unique identifier as key and its opinion as value, named \textit{store}.
\end{itemize}

The first function in the algorithm \textit{addValidator} is to be invoked by the coordinator whenever a new validator joins the proposed system. This function takes the address as an argument and verifies if the validator paid the fee or not. If paid, then the validator's public key is added to \textit{valid\_validators}. The second function \textit{setup} is restricted to the coordinator, as it is used to initialize the \textit{store}. The third function \textit{get\_opinion} is also invoked by the coordinator alone. This function basically returns the store mapping to the coordinator. This mapping is used to calculate the trust of the validator. The last function named \textit{set\_opinion} is the function that is invoked by both validators and coordinator. This function is used to populate the store mapping. It takes two arguments, namely \textit{key} and \textit{opinion}. The argument \textit{key} corresponds to the unique identifier of a particular validator, and \textit{opinion} corresponds to the validator's opinion on the given set of weights.

\section{Implementation}
\subsection{Technologies Used}
\subsubsection{Foundry:}
Foundry \cite{r24} is an open source Rust based smart contract development tool kit that is used for developing, testing and deploying Smart Contracts and other decentralized applications to Ethereum blockchain. This toolkit leverages on various key properties of Rust like being statically typed and compiled, thus gaining in terms of performance and resources. This toolkit supports "fuzzing", fuzzing is the process of an automated testing process that tests various edge cases and invalid input. Foundry is a combination of the following tools:
\begin{itemize}
    \item Anvil: It can be used for setting up a local blockchain network that can be used while developing the smart contract
    \item Cast : It is a set of CLI commands that can be used to interact with the smart contract like sending transactions and read data from the network. It provides a set of sub commands that can be used for debugging smart contracts.
    \item Forge: It is a set of CLI commands used for testing, developing and deploying of the smart contract.
\end{itemize}

\subsubsection{Python Socket.IO}
Python Socket.io \cite{r25} is a python implementation of the Socket.IO protocol. Socket.IO protocol facilitates real-time, bidirectional and event based communication between a client and server. This protocol is based on the Web Socket protocol, with HTTP polling as a fallback option. This library is used to establish the communication between the node, validator, and coordinator. The proposed solution leverages on the events based communication python Socket.io provides.
\subsubsection{Web3 Py}
Web3 py\cite{r26} is a python library that enables developers to interact with Ethereum blockchain. It is widely used in decentralized applications to initiate transactions, interact with smart contracts and for reading and writing data to the blockchain. One of the key features, the ability to interact with smart contracts using Solidity. Web3.py was derived directly from the Web3.js JavaScript API, but is centered more around the needs of Python developers. With Web3py, the developers can integrate Ethereum functionality into their code, thus building a decentralized application.

\subsection{Implementation Canvas}
Nodes and Validators are connected using the web socket connection. This communication was implemented using python-socketio. This type of communication is event based. Weights of the trained model are encrypted using PGP wrapper for python\cite{r32}, the cipher is then uploaded to IPFS using ipfshttpclient \cite{r33}. The default encryption algorithm used for this process is RSA-2048. The obtained CID is then sent to the endpoint in the receiver, where the reverse of the above-mentioned process takes place. Smart Contract used in the proposed method is invoked by the address and the functional interaction are done by the web3.py module.
\section{Distinguishing Features of Proposed Model}
\subsection{Transparent decision-making process}
Blockchain, as mentioned in the previous section, serves as a distributed and immutable ledger where the record of the decision-making process is maintained. This record includes the individual opinion of the validator on models submitted by the other validator. The availability of the data on the blockchain ensures the authenticity of the information, enhancing the transparency in the decision-making process. In scenarios where the number of accepted weights is less than $\zeta$, the data stored in the blockchain is used to create the global model.
\subsection{Resilient against attacks}
As mentioned in the previous section, the IPFS in conjunction with Encryption prevents Man in the middle attack. Since IPFS is a distributed file system, there is no single point of failure. This combination reduces the centralization of the whole system. Since only the CID is only communicated between the nodes, it significantly reduces the inter-nodal communication. This streamlined approach optimizes efficiency, leading to a highly robust network architecture.
\subsection{Hierarchical Structure}
The arrangement of the nodes in a hierarchy provides advantages which include reduction in impact of attacks and improves the scalability. Validators requirement to pay a participation fee adds an extra layer of trustworthiness to the system, as this fee acts as a guarantee of their credibility. Furthermore, the implementation of the federated algorithm in two level enhances the robustness.
\subsection{Validation of Weights}
The combination of the Novelty Detection and Snowball protocol ensures that the global model is generated by the  the weights that are accepted by all the validators, thus making the proposed model resilient against the data poisioning attacks. As snowball protocol is scalable and need a constant communication cost, there is minimum overhead. This enforces creation of the better model that would be resilient against attacks. Furthermore, in absence of the trust the influence based final model generation ensures the model to be learned from the most trusted validator.

\section{Results}
\subsection{Setup}
For testing the performance of the proposed model, we have conducted experiments on a system with following configurations:
\begin{itemize}
    \item AMD Ryzen 5 4600H with Radeon Graphics @ 3.0GHz
    \item 16 GB of memory with Pop! OS 22.04 LTS 
\end{itemize}
The proposed model's performance evaluation utilized three datasets: Mushroom Dataset \cite{r27}, Pima Indians Diabetes Database \cite{r29}, and Thyroid Disease Data Set \cite{r28}. These datasets were partitioned into test and train sets, with the model trained on the train data and validated against the test set. Test accuracy served as the primary metric for evaluation. Due to the inherent imbalance in the Pima Indians Diabetes Database and Thyroid Disease Data Set, class weights were employed during the model generation process to address this imbalance. Logistic Regression and SGDClassifier models, were considered for this experiment. Each experiment is repeated 10 times to minimize randomness.

All experiments were carried out with 5 validators,with its federation consisting of  2 nodes. The Snowball protocol's parameters $\alpha$,$\beta$,k are set to 3, 3, and 4, respectively, on the other hand, the \textit{Minimal Consensus Index} is varied from 0 to 5.

To simulate label flipping, two new metrics were introduced:
\begin{itemize}
    \item Flip: This parameter indicates the number of validators with malicious data in the entire network. 
    \item Proportion: This parameter represents the fraction of labels flipped, ranging from 0 to 8.
\end{itemize}

If Flip, Proportion were set to 2,4 respectively, 2 validators out of 5 were generating 40 \% of the label flipped data. 
The obtained results were compared with the Flower framework.
\subsection{Observations}
\begin{figure}
    \centering
    \begin{subfigure}[t]{0.35\textwidth}
    \includegraphics[width=\textwidth]{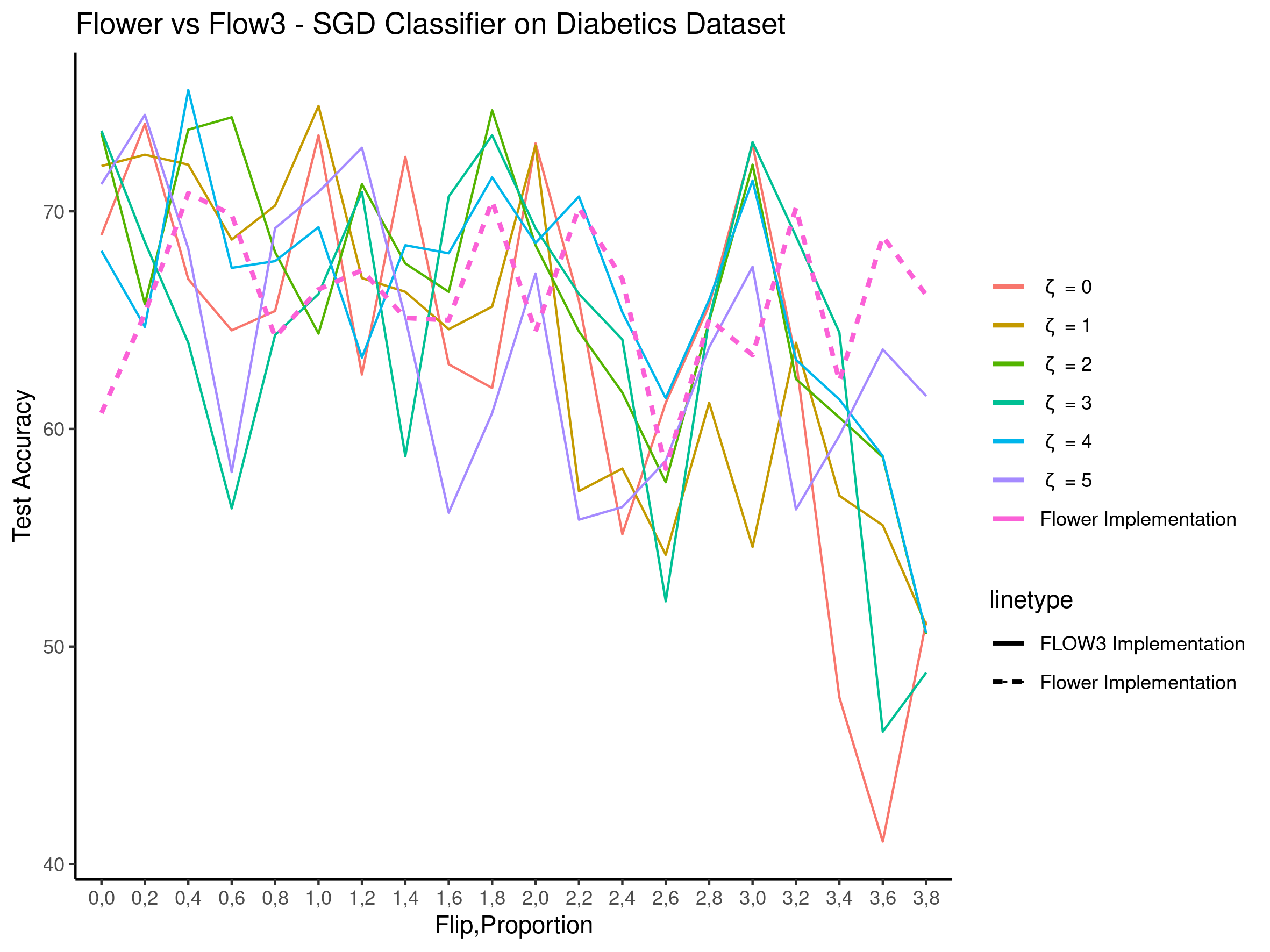}
    \end{subfigure}
    \begin{subfigure}[t]{0.35\textwidth}
        \includegraphics[width=\textwidth]{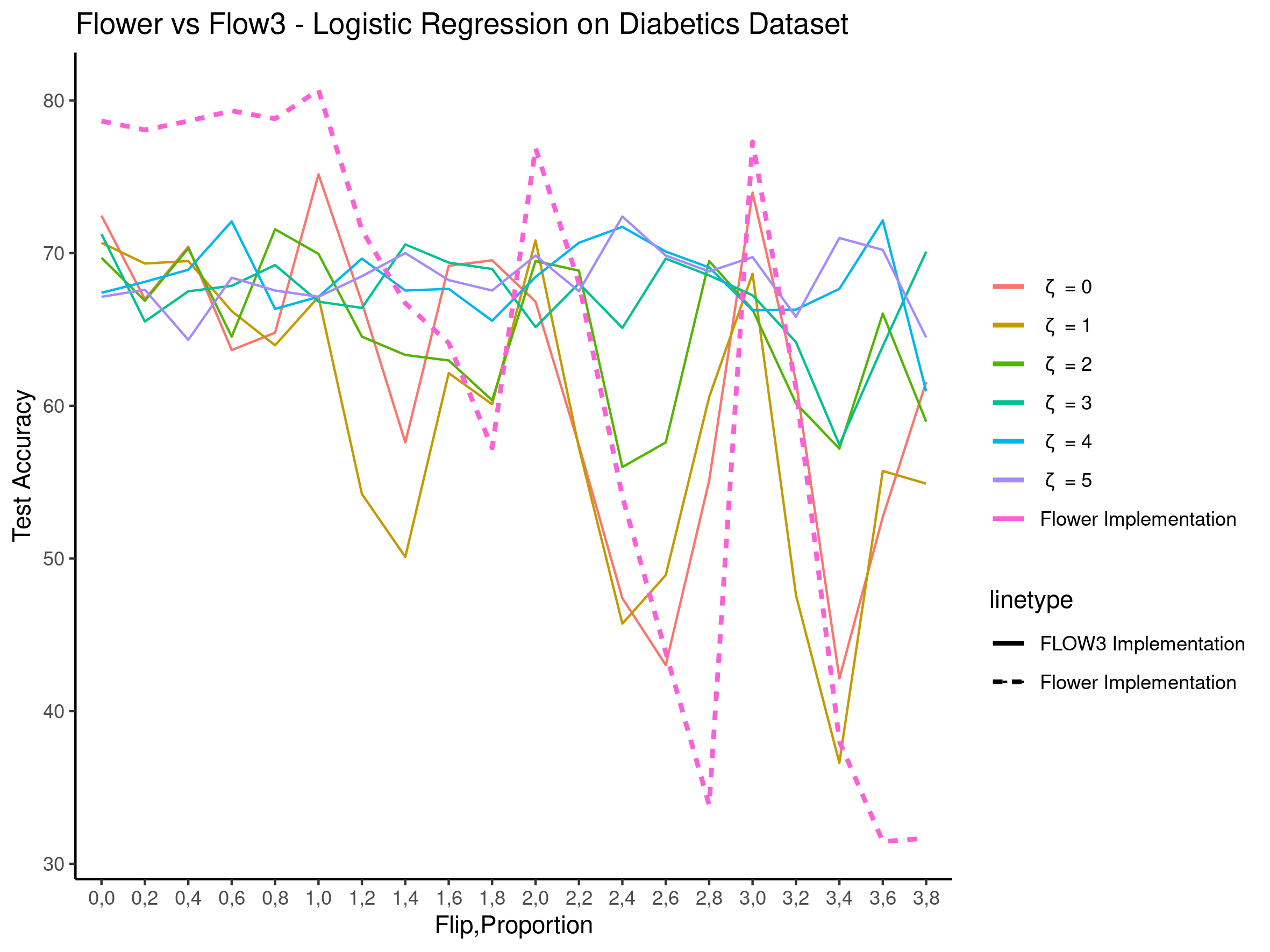}
    \end{subfigure}
    \caption{Evaluation on Diabetic Dataset}
    \label{diabetic}

    \centering
    \begin{subfigure}[t]{0.35\textwidth}
    \includegraphics[width=\textwidth]{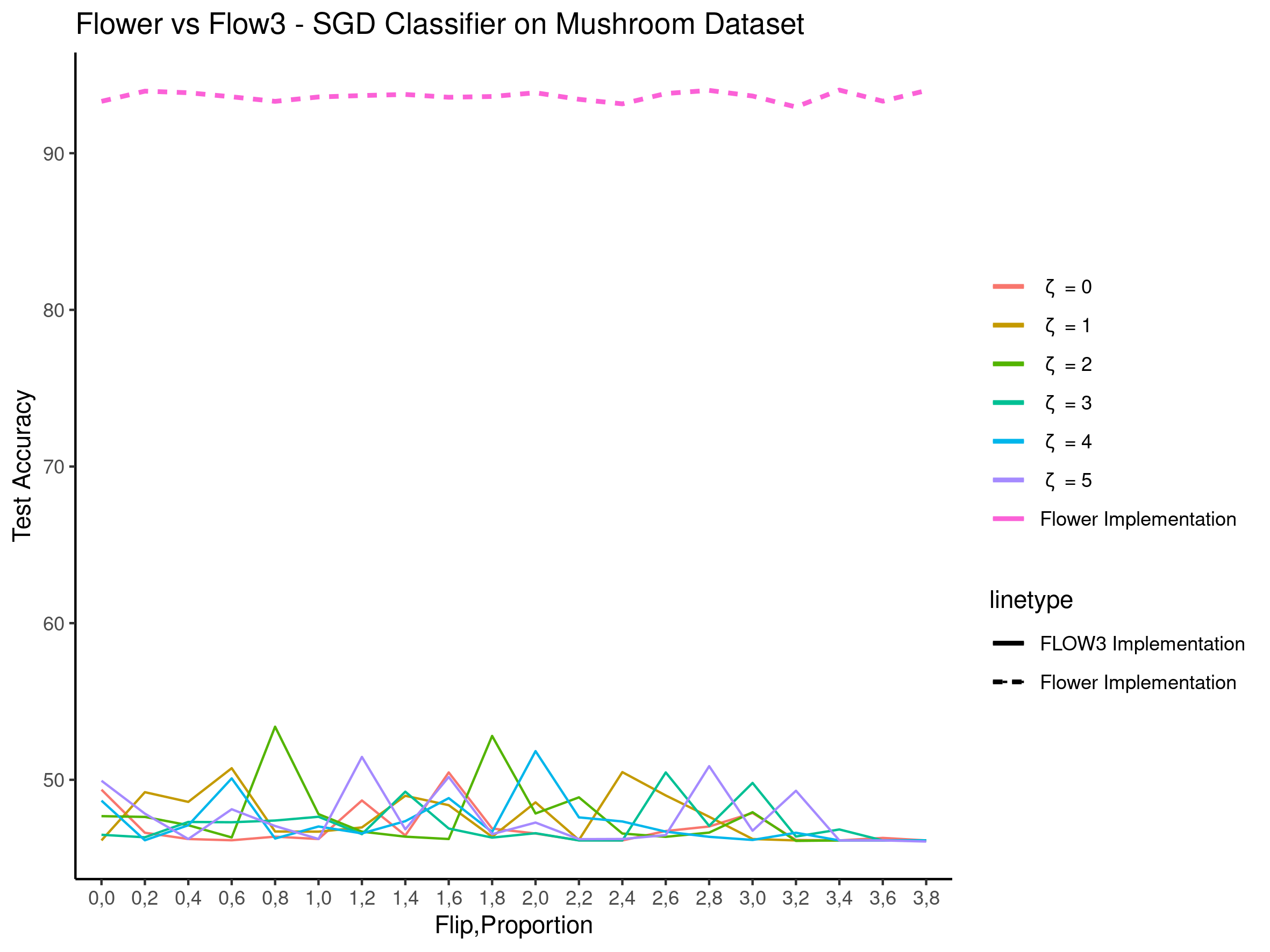}
    \end{subfigure}
    \begin{subfigure}[t]{0.35\textwidth}
        \includegraphics[width=\textwidth]{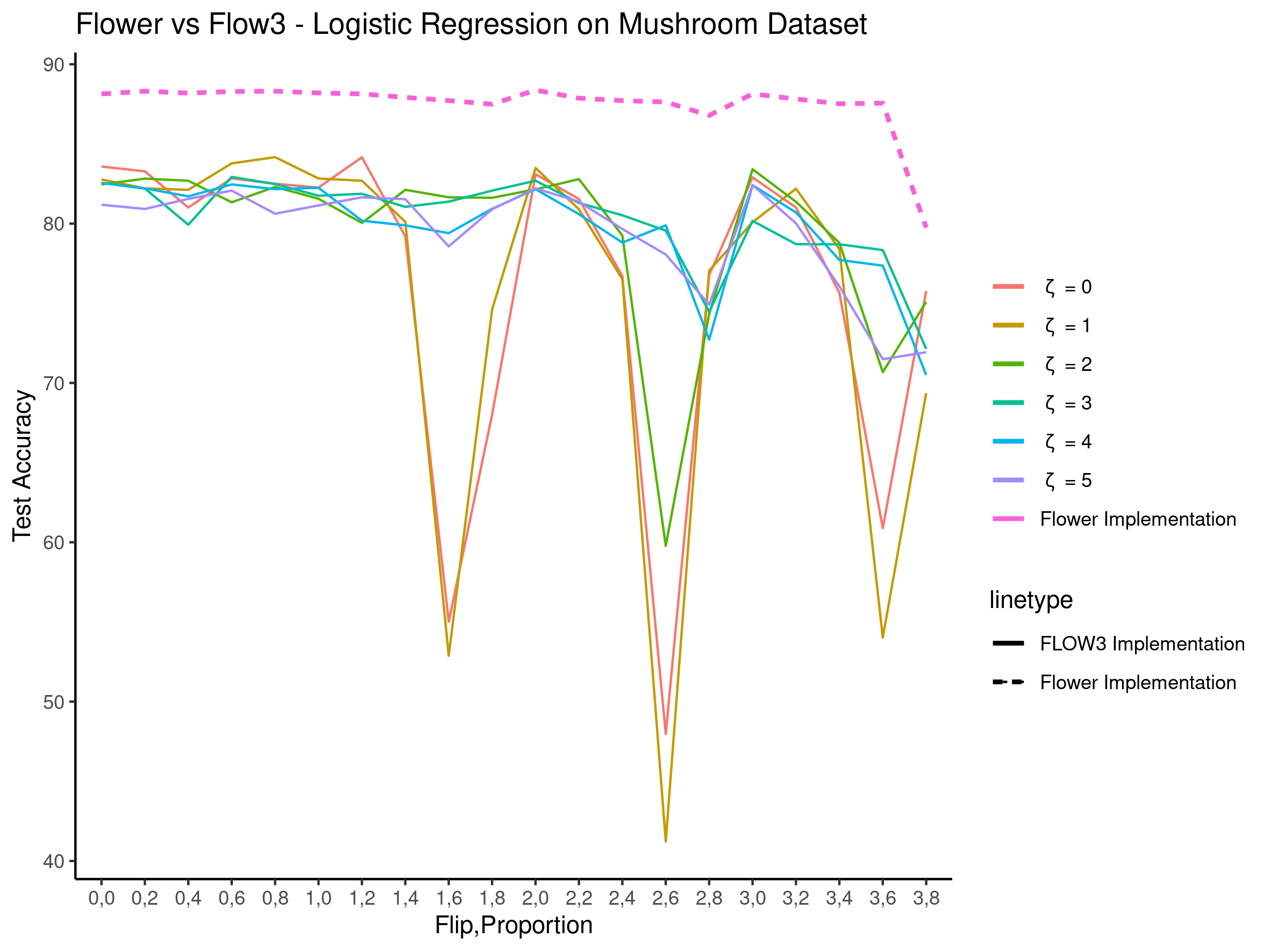}
    \end{subfigure}
    \caption{Evaluation on Mushroom Dataset}
    \label{mushroom}
    \centering
    \begin{subfigure}[t]{0.35\textwidth}
    \includegraphics[width=\textwidth]{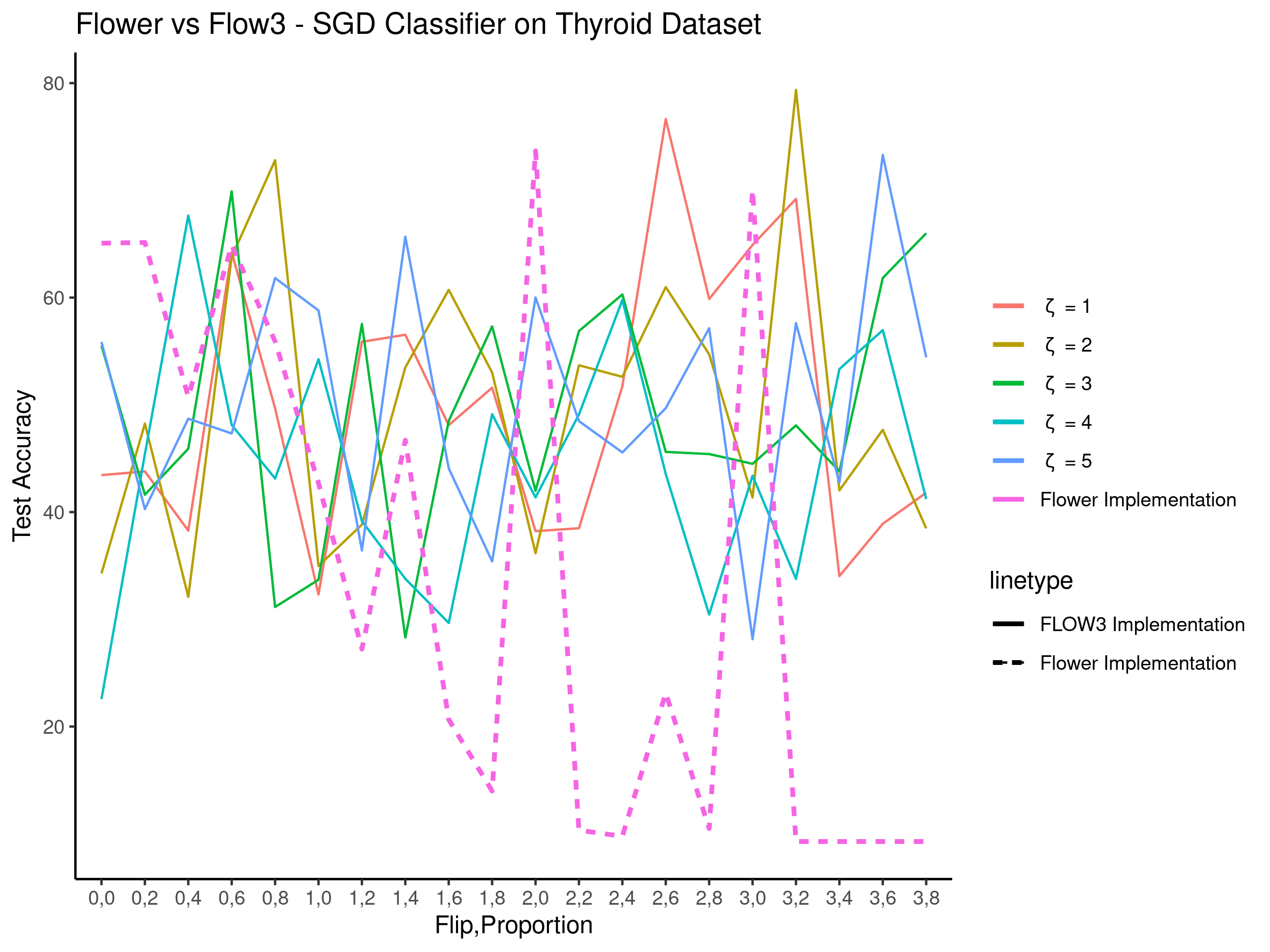}
    \end{subfigure}
    \begin{subfigure}[t]{0.35\textwidth}
        \includegraphics[width=\textwidth]{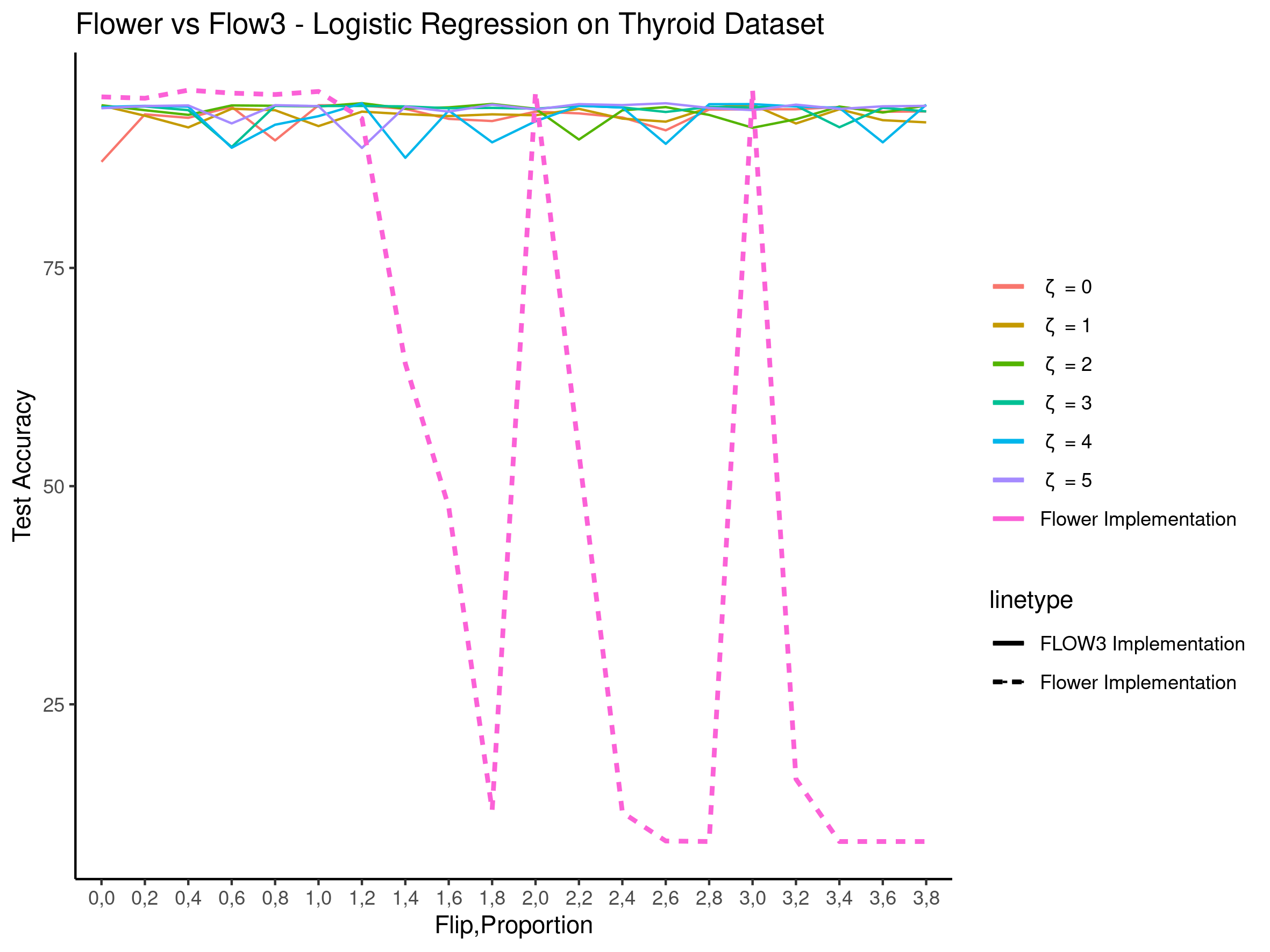}
    \end{subfigure}
    \caption{Evaluation on Thyroid Dataset}
    \label{thyroid}
\end{figure}
The dotted line in the plots signifies the test accuracy of the Flower-generated models. Generally, as the magnitude of the flip and proportion increases, the overall model performance decreases. Specifically, for lower $\zeta$ values, the models generated by the proposed platform exhibit significant oscillations.
However, for higher $\zeta$ values, the proposed platform's models remain stable under various attacks, albeit with a decrease in accuracy. Overall, it can be concluded that the performance of the proposed platform is comparable to Flower, except for a few specific cases.

In the Diabetics dataset (Figure \ref{diabetic}), for Logistic Regression, the accuracy of the proposed platform's model remains stable compared to the Flower-generated models. While Flower models perform well in the absence of attacks, their accuracy declines during attacks, especially with lower $\zeta$ values, leading to significant oscillations. Regarding SGDClassifier, the models generated by the proposed platform outperform Flower models in most cases. However, in extreme scenarios where the flip parameter is set to 3, the Flower-generated models surpass the proposed platform's performance.

In the Mushroom dataset (Figure \ref{mushroom}), for Logistic Regression, the Flower implementation outperforms the proposed model, albeit by a small margin. Notably, when $\zeta$ is set to 0 and 1, the test accuracy of the models fluctuates considerably. In the case of SGD Classifier, models generated using Flower significantly outperforms the models generated by the proposed platform.

In the Thyroid dataset (Figure \ref{thyroid}), concerning Logistic Regression, the accuracy of the proposed platform's model remains stable, unlike the Flower-generated models. Flower models perform well in the absence of attacks, but their accuracy declines during attacks. Regarding SGD Classifier, the performance of the models generated by the platform is comparable to those created using Flower. Notably, the proposed platform's models exhibit relative stability in accuracy, whereas Flower models tend to oscillate.

\section{Conclusion and Future Work}
In this paper, we discussed a new system to implement Federated learning in a decentralized manner. In the proposed system, the nodes in the network are arranged in a tree shaped configuration. Novelty detection in conjunction with Snowball protocol is used to create a scalable and robust federated learning setting.  Blockchain is used as a record keeper which maintains the record of decision-making process, thus bring transparency. By the design choices made, the proposed system is robust against Man in the middle attack, Data Poisoning and Model Poisoning attacks.

Further work can be done in the following lines, implementing complex machine learning algorithms like Neural Networks, CNNs, LSTMs, implementing consensus protocols like RAFT, leveraging on highly scalable networks like Avalanche, Polkadot instead of Ethereum. Usage of better trust and influence calculation.

\bmhead{Acknowledgments}
The authors dedicate this work to Bhagwan Sri Sathya Sai Baba, the Founder Chancellor of SSSIHL, for being a constant source of inspiration. 
\section*{Declarations}
\begin{itemize}
\item Funding: Not Applicable
\item Conflict of Interest:On behalf of all authors, the corresponding author states that there is no conflict of interest
\item Ethical standards This article does not contain any studies with human
participants or animals performed by any of the authors
\item Informed Consent: Not Applicable
\item Author contribution: VRK contributed to the final version of the manuscript. PKB supervised the project.
\item Code Availability: Code will be available publicly in github after
acceptance.

\end{itemize}

\bibliography{sn-bibliography}

\end{document}